\documentclass{appolb}
\usepackage{graphicx}

\begin{document}
\title{Magnetic Anisotropy in La$_{0.8}$Sr$_{0.2}$MnO$_3$: Electron Spin Resonance
\thanks{Presented at the Strongly Correlated Electron Systems
Conference, Krak\'ow 2002}%
}


\author{J. Deisenhofer$^{1}$\thanks{Corresponding author: joachim.deisenhofer@physik.uni-augsburg.de}, H.-A.
Krug von Nidda$^{1}$, A. Loidl$^{1}$, M.V. Eremin$^{2}$, V.A.
Ivanshin$^{2}$, T.~Kimura$^{3}$ and Y. Tokura$^{3}$
\address{$^{1}$Experimentalphysik~V, EKM, Institut
f\"{u}r Physik, Universit\"{a}t Augsburg, D-86135 Augsburg,
Germany\\ $^{2}$Kazan State University, 420008 Kazan, Russia\\
$^{3}$ Department of Applied Physics, University of Tokyo, Tokyo
113-0033, Japan} } \maketitle


\begin{abstract}
We report on Ferromagnetic-Resonance experiments in a single
crystal of La$_{0.8}$Sr$_{0.2}$MnO$_3$ in the temperature range
from 4 to 300 K. The observed anisotropy of the resonance line
changes on crossing the transition from the orthorhombic $O$-phase
to the rhombohedral $R$-phase at $T\approx 100$ K and indicates a
reorientation of the spins at about 130 K.

\end{abstract}




\section{Introduction}

The importance of magnetic inhomogeneities in understanding the
complex phase diagrams of manganite systems like
La$_{1-x}$Sr$_{x}$MnO$_3$ \cite{Phasedia} becomes more and more
evident, only to mention phase separation scenarios, chemical
inhomogeneities or the recent description of the Colossal Magneto
Resistance effect as a Griffith phase \cite{Inhoms}. Electron Spin
Resonance not only proves a very efficient tool in investigating
phenomena like orbital ordering or inhomogeneities in the
paramagnetic phase \cite{EPR}, but it is also very sensitive to
changes in the magnetization behavior in the magnetically ordered
regime \cite{Hysteresis,Viglin01,HighfieldESR}.

The details of the experimental setup and measurement procedures
are described e.g.~in \cite{EPR}. The single crystal of
La$_{0.8}$Sr$_{0.2}$MnO$_3$ has been grown by the floating zone
method described by Urushibara and coworkers \cite{Phasedia}. A
thin disc parallel to the (001)-plane has been prepared in order
to control demagnetization effects. We measured at X-Band
frequency (9.4 GHz) and rotated the sample with the external
magnetic field $\mathbf{H}$ applied within the disc's plane (see
inset of Fig.~\ref{Hresvsang}(c)).

\section{Experimental results and discussion}

\begin{figure}[!ht]
\begin{center}
\includegraphics[width=0.6\textwidth,clip]{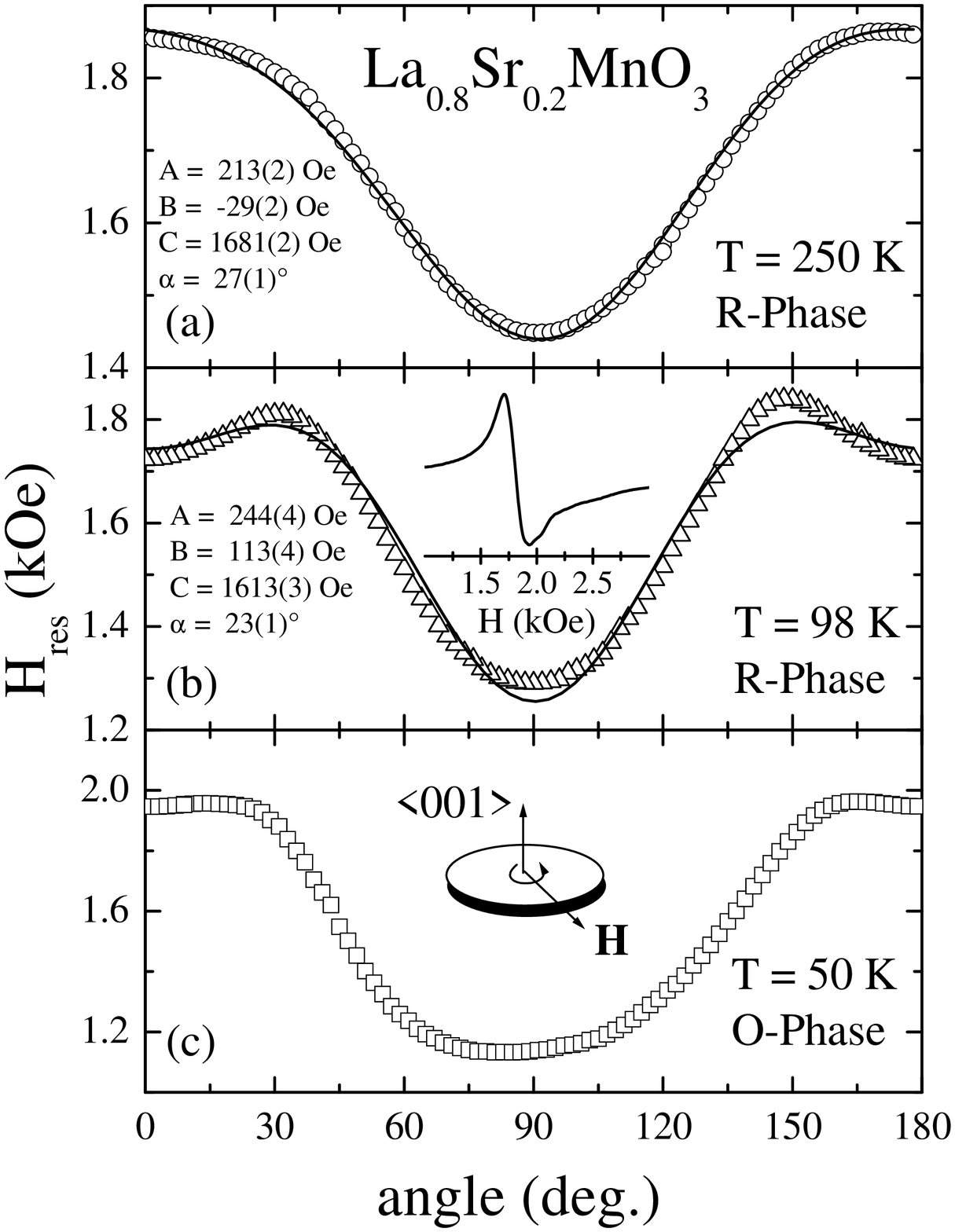}
\end{center}
\caption{Angular dependence of $H_{\mathrm{res}}$ at (a) 250 K,
(b) 98 K and (c) 50 K revealing changes in symmetry with
temperatures. Inset of (b) shows the field derivation of the
Ferromagnetic-Resonance absorption. The solid lines are fits using
equation (1).} \label{Hresvsang}
\end{figure}

The resonance spectra (a typical one is shown in the inset of
Fig.~\ref{Hresvsang}(b)) reveal distortions of the main resonance
above the resonance field similar to the ones observed in
\cite{Viglin01}. We fitted the main resonance line with a
Lorentzian lineshape and observed an anisotropy in both the
linewidth and the resonance field $H_{\mathrm{res}}$. The latter
is shown in Fig.~\ref{Hresvsang} for three different temperatures:
At about 95 K (upon heating) the system changes from orthorhombic
($O$-Phase) to rhombohedral symmetry ($R$-Phase). The transition
to the paramagnetic state takes place at $T_{\mathrm C}\approx
310$ K. In the $O$-phase at 50 K a twofold symmetry  with an
asymmetric broad minimum is found, which can be attributed to the
deviation of the orthorhombic axes from cubic symmetry.

Right above the structural transition at 98 K we observe an
anisotropy which we describe by a superposition of a fourfold and
a twofold symmetry using
\begin{equation}
H_{\mathbf{res}}(\phi)=A\sin(2\phi) + B\sin(4(\phi-\alpha)) + C,
\end{equation}
where $\phi$ denotes the rotation angle in the (001)-plane. Far
above the transition at 250 K again a twofold symmetry is found,
indicating a uniaxial magnetocrystalline anisotropy. The fit with
eq.~(1), however, includes still a fourfold contribution. The fits
and fitparameters are shown in the corresponding panels of Fig.~1.
Besides the fact that the dominant uniaxial anisotropy parameter
$A$ decreases, the fourfold parameter $B$ even becomes negative,
indicating a change of hard and easy axes between 98 K and 250 K
within the $R$-phase. A fourfold symmetry has also been reported
by Viglin \textit{et al.}~\cite{Viglin01} yielding anisotropy
fields of $H_{A_1}\approx 130$ Oe by fitting with the usual
angular dependence of cubic symmetry in the cube face plane
\cite{Gurevich}. The low value of $C\approx 1.6$ kOe in contrast
to the resonance observed in a sphere \cite{Viglin01} can be
explained by taking into account the demagnetization effects for a
disc \cite{Gurevich}. Moreover, Szymczak \textit{et
al.}~\cite{Hysteresis} report on FMR and magnetization
measurements in the (110)-plane revealing an uniaxial
magnetocrystalline anisotropy and a change of the easy and hard
directions concomitantly to the structural transition.

\begin{figure}[!hb]
\begin{center}
\includegraphics[width=0.45\textwidth,clip, angle=-90]{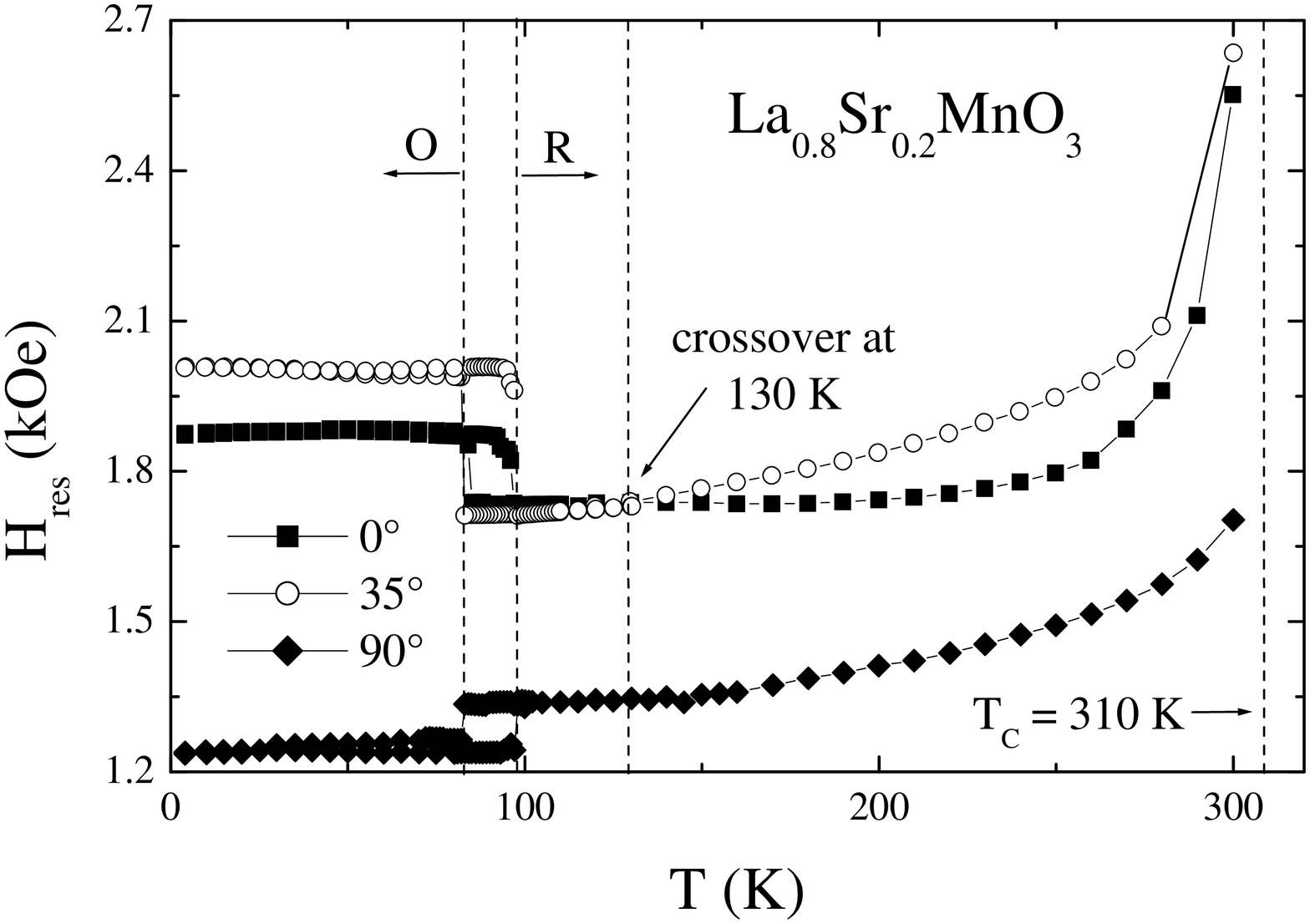}
\end{center}
\caption{Temperature dependence of $H_{\mathrm{res}}$ with the
external magnetic field applied parallel to the directions of the
three extrema from Fig.~1.} \label{HresVsT}
\end{figure}

Therefore we measured the temperature dependence of the resonance
line with the external field parallel to the three extrema at
0$^\circ$, 35$^\circ$ and 90$^\circ$ determined from the angular
dependence in Fig.~\ref{Hresvsang}(b). The temperature dependence
of $H_{\mathrm{res}}$ for all three directions is shown in
Fig.~\ref{HresVsT}: The anisotropy in the $O$-phase is almost
constant for $T<$ 80 K, whereas for 80 K $<T<$ 95 K a hysteretical
behavior is observed, which reportedly accompanies the structural
phase transition from orthorhombic to rhombohedral symmetry
\cite{Hysteresis}. The $R$-phase is characterized by a decrease in
anisotropy and only within the narrow temperature range of 95 K
$<T<$ 130 K the angular dependence exhibits the three extrema as
seen in Fig.~\ref{Hresvsang}(b). For $T> 130$ K the angular
dependence reveals a twofold symmetry as in
Fig.~\ref{Hresvsang}(a). In the temperature dependence this change
is indicated at the crossover of the data for 0$^\circ$ and
35$^\circ$. Towards the magnetic transiton a shift of the
resonance to higher fields is observed for all three directions,
which is due to the decrease of the magnetization on approaching
$T_{\mathrm C}$.

\section{Conclusion}

In conclusion, we found that the anisotropy in the ferromagnetic
(001)-plane can be interpreted as a superposition of a cubic
fourfold symmetry and a twofold symmetry due to anisotropic
ferromagnetic superexchange interactions like in LaMnO$_3$. The
fourfold contribution is strongest just above the structural
transition, where the system tends to restore cubic symmetry. The
spin-reorientation is found to take place at 130 K far above the
structural transition.

We are grateful to V. Tsurkan for help with the preparation of the
disc. This work was supported in part by the BMBF under contract
no. 13N6917 (EKM), by the Deutsche Forschungsgemeinschaft (DFG)
via the SFB 484 and DFG-project No.~436-RUS 113/566/0, and by
INTAS (project no 97-30850). The work of M.V.E was partially
supported by RFBR grant no. 00-02-17597 and NIOKR Tatarstan.
M.V.E.~und V.A.I.~were supported in part by the Swiss National
Science Foundation (grant no.~7SUPJ062258).

\end{document}